\documentclass[aps,prl,twocolumn,groupedaddress]{revtex4}
\usepackage{graphicx}
\def\vc#1{\mbox{\boldmath $#1$}}

\begin{document}

%\preprint{}

\title{Alpha-particle condensation in $^{16}$O via a full four-body OCM 
calculation}


\author{Y.~\textsc{Funaki}$^1$, T.~\textsc{Yamada}$^2$, H.~\textsc{Horiuchi}$^3$, G.~\textsc{R\"opke}$^4$, P.~\textsc{Schuck}$^{5,6}$ and A.~\textsc{Tohsaki}$^3$}
%\thanks{}
\affiliation{$^1$Nishina Center for Accelerator-Based Science, The Institute of Physical and Chemical Research (RIKEN), Wako 351-0198, Japan}
\affiliation{$^2$Laboratory of Physics, Kanto Gakuin University, Yokohama 236-8501, Japan}
\affiliation{$^3$Research Center for Nuclear Physics (RCNP), Osaka University, Osaka 567-0047, Japan}
\affiliation{$^4$Institut f\"ur Physik, Universit\"at Rostock, D-18051 Rostock, Germany}
\affiliation{$^5$Institut de Physique Nucl\'eaire, CNRS, UMR 8608, Orsay, F-91406, France}
\affiliation{$^6$Universit\'e Paris-Sud, Orsay, F-91505, France} 


\date{\today}

\begin{abstract}
In order to explore the $4\alpha$-particle condensate state in $^{16}$O, we solve a full four-body equation of motion based on the $4\alpha$ OCM (Orthogonality Condition Model) in a large $4\alpha$ model space spanned by Gaussian basis functions. A full spectrum up to the $0_6^+$ state is reproduced consistently with the lowest six $0^+$ states of the experimental spectrum. The $0^+_6$ state is obtained at about $2$ MeV above the $4\alpha$ breakup threshold and has a dilute density structure, with a radius of about 5 fm. The state has an appreciably large $\alpha$ condensate fraction of $61$ \%, and a large component of $\alpha+^{12}$C$(0_2^+)$ configuration, both features being reliable evidence for this state to be of $4\alpha$ condensate nature. 
\end{abstract}

%\keywords{}

\maketitle

 It is well established that $\alpha$-clustering plays a very important role 
for the structure of lighter nuclei~\cite{tang,carbon}. The importance of 
$\alpha$-cluster formation also has been discussed in infinite nuclear 
matter, where $\alpha$-particle type condensation is expected at low density~\cite{roepke}, quite in analogy to the recently realised Bose-Einstein 
condensation of bosonic atoms in magneto-optical traps~\cite{stringari}. On the other hand, for trapped fermions, quartet condensation also is an emerging 
subject, discussed, so far, only theoretically~\cite{lecheminant}. In nuclei the bosonic constituents always are only very few in number, nevertheless possibly giving rise to clear condensation characteristics, as is well known from 
nuclear pairing~\cite{ring}. Concerning $\alpha$-particle condensation, only the Hoyle state, i.e. the $0_2^+$ state in $^{12}$C has clearly been established, so far. Several papers of the past~\cite{horiuchi_ocm,kamimura,uegaki,baye} and also more recently~\cite{thsr,funaki1,neff} have by now established beyond any doubt that the Hoyle state, only having about one third of saturation density, can be described, to good approximation, as a product state of three $\alpha$-particles, condensed, with their c.o.m. motion, into the lowest mean field $0S$-orbit~\cite{matsumura,yamada_ocm}.

 The establishment of this novel aspect of the Hoyle state naturally 
leads us to the speculation about $4\alpha$-particle condensation in 
$^{16}$O, which is the focus of this work. The $0^+$ spectrum of $^{16}$O has, in the past, very well been reproduced up to about $13$ MeV excitation energy, including 
the ground state, with a semi-microscopic cluster model, i.e. the  
$\alpha + ^{12}$C OCM (Orthogonality Condition Model)~\cite{Suz76}. 
In particular, this model calculation, as well as that of an $\alpha+^{12}$C Generator-Coordinate-Method one~\cite{baye2}, demonstrates that the $0_2^+$ 
state at $6.05$ MeV and the $0_3^+$ state at $12.05$ MeV have 
$\alpha + ^{12}$C structures~\cite{Hor68} where the $\alpha$-particle 
orbits around the $^{12}$C$(0_1^+)$-core in an $S$-wave and around the $^{12}$C$(2_1^+)$-core in a $D$-wave, respectively. Consistent results were later obtained by 
the $4\alpha$ OCM calculation within the harmonic oscillator 
basis~\cite{Kat92}. However, the model space adopted in 
Refs.~\cite{Suz76,Kat92,baye2} is not sufficient to account simultaneously for the $\alpha+ ^{12}$C and the $4\alpha$ 
gas-like configurations. On the other hand, the $4\alpha$-particle condensate 
state was first investigated in Ref.~\cite{thsr} and its existence was 
predicted around the $4\alpha$ threshold with a new type of microscopic wave 
function of $\alpha$-particle condensate character. While that wave function 
can well describe the dilute $\alpha$ cluster states as well as shell-model-like states, other structures such as $\alpha + ^{12}$C clustering 
are smeared out and only incorporated in an average way. Since there exists 
no calculation, so far, which reproduces both the $4\alpha$ gas and 
$\alpha+^{12}$C cluster structures simultaneously, it is crucial to perform an extended calculation for the simultaneous reproduction of both kinds of structures, which will give a decisive benchmark for the existence 
of the $4\alpha$-particle condensate state from a theoretical point of view.

 The purpose of this Letter is to explore the $4\alpha$ condensate state by 
solving a full OCM four-body equation of motion without any assumption with 
respect to the structure of the $4\alpha$ system. Here we take the $4\alpha$ 
OCM with Gaussian basis functions, the model space of which is large enough 
to cover the $4\alpha$ gas, the  $\alpha +^{12}$C cluster, as well as the 
shell-model configurations. The OCM is extensively described in Ref.~\cite{saitoh}. Many successful applications of OCM are reported in Ref.~\cite{carbon}.
The $4\alpha$ OCM Hamiltonian is given as follows:

\begin{eqnarray}
&&\hspace{-0.7cm}{\cal H}=\sum_{i}^{4}T_i - T_{\rm cm}+ \sum_{i<j}^4
\Big[ V_{2\alpha}^{({\rm N})}(i,j)+V^{({\rm  C})}_{2\alpha}(i,j) \nonumber \\
&&\hspace{-0.7cm} + V_{2\alpha}^{({\rm P})}(i,j) \Big]  +\sum_{i<j<k}^4 
V_{3\alpha}(i,j,k)+ V_{4\alpha}(1,2,3,4), \label{eq:hamil}
\end{eqnarray}
where $T_i$, $V_{2\alpha}^{({\rm N})}(i,j)$, $V_{2\alpha}^{({\rm C})}(i,j)$, 
$V_{3\alpha}(i,j,k)$ and $V_{4\alpha}(1,2,3,4)$ stand for the operators of 
kinetic energy for the $i$-th $\alpha$ particle, two-body, Coulomb, three-body 
and four-body forces between $\alpha$ particles, respectively. The center-of-mass kinetic energy $T_{\rm cm}$ is subtracted from the Hamiltonian. 
$V_{2\alpha}^{({\rm P})}(i,j)$ is the Pauli exclusion operator~\cite{kukulin}, by which 
the Pauli forbidden states between two $\alpha$-particles in $0S$, $0D$ and $1S$ 
states are eliminated, so that the ground state with the shell-model-like 
configuration can be described correctly. The effective $\alpha$-$\alpha$ interaction $V_{2\alpha}^{\rm (N)}$ is constructed 
by the folding procedure from two kinds of effective two-nucleon forces. One is 
the Modified Hasegawa-Nagata (MHN) force~\cite{mhn} and the other is the 
Schmidt-Wildermuth (SW) force~\cite{sw}, see Refs.~\cite{yamada_ocm} and ~\cite{kurokawa} for applications, respectively. We should note 
that the folded $\alpha$-$\alpha$ potentials reproduce the $\alpha$-$\alpha$ 
scattering phase shifts and energies of the $^8$Be ground state and of the Hoyle state. The three-body force is phenomenologically introduced so as to fit the ground state energy of the $^{12}$C. The same force parameter set as used in Refs.~\cite{yamada_ocm} or ~\cite{kurokawa} is adopted in the present calculation. In addition, the phenomenological four-body force is adjusted to the ground state energy of $^{16}$O. The three-body and four-body forces are short-range, and, hence, they only act in compact configurations.

 Utilizing the Gaussian expansion method~\cite{GEM} for the choice of 
variational basis functions, the total wave function $\Psi$ of the $4\alpha$ 
system is expanded in terms of Gaussian basis functions as follows: 
\begin{eqnarray}
&&\hspace{-0.7cm} \Psi(0_n^+)=\sum_{c, \nu} A^n_{c}(\nu)\Phi_{c}(\nu), \\
&&\hspace{-0.7cm} \Phi_{c}(\nu) ={\cal \widehat S} \Big[[\varphi_{l_1}
(\vc{r}_1,\nu_1)\varphi_{l_2}(\vc{r}_2,\nu_2)]_{l_{12}}  \varphi_{l_3}
(\vc{r}_3,\nu_3) \Big]_{J}, \label{eq:3}
\end{eqnarray}
where $\vc{r}_1$, $\vc{r}_2$ and $\vc{r}_3$ are the Jacobi coordinates 
describing internal motions of the $4\alpha$ system. ${\cal \widehat S}$ 
stands for the symmetrization operator acting on all $\alpha$ particles 
obeying Bose statistics. $\nu$ denotes the set of size parameters $\nu_1,
\nu_2$ and $\nu_3$ of the normalized Gaussian function, $\varphi_{l}(\vc{r},\nu_i)=
N_{l,\nu_i}r^l\exp{(-\nu_i r^2)} Y_{l m}(\hat{\vc r})$, and $c$ the set of 
relative orbital angular momentum channels $[[l_1,l_2]_{l_{12}},l_3]_J$ 
depending on either of  the coordinate type of $K$ or $H$~\cite{GEM}, 
where $l_1$, $l_2$ and $l_3$ are the orbital angular momenta with respect 
to the corresponding Jacobi coordinates. The coefficients $A^n_{c}(\nu)$ 
are determined according to the Rayleigh-Ritz variational principle. 
\begin{figure}[htbp]
\begin{center}
\includegraphics[width=7.1cm]{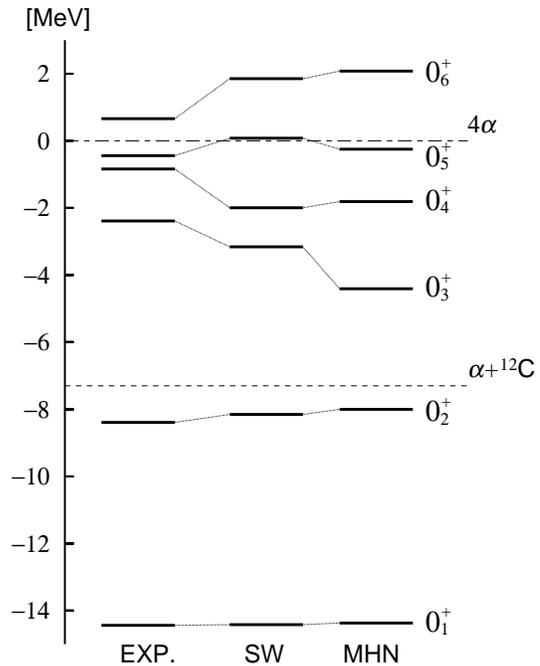}
\caption{Comparison of energy spectra between experiment and the present calculation. Two kinds of effective two-body nucleon-nucleon forces MHN and SW are adopted (see text). Dotted and dash-dotted lines denote the $\alpha + ^{12}$C and $4\alpha$ thresholds, respectively. Experimental data are taken from Ref.~\cite{ajze}, and from Ref.~\cite{wakasa} for the $0_4^+$ state. The assignments with experiment are tentative, see, however, detailed discussion in the text.}\label{fig:1}
\end{center}
\end{figure}

 Figure~\ref{fig:1} shows the energy spectrum with $J^\pi=0^+$, which is obtained by diagonalizing the Hamiltonian, Eq.~(\ref{eq:hamil}), in a model space as large as given by 5120 Gaussian basis functions, Eq.~(\ref{eq:3}). It is confirmed that all levels are well converged. With the above mentioned effective $\alpha$-$\alpha$ forces, we can reproduce the 
full spectrum of $0^+$ states, and tentatively make a one-to-one 
correspondence of those states with the six lowest $0^+$ states of the 
experimental spectrum. In view of the complexity of the situation, the 
agreement is considered to be very satisfactory. 
\begin{table}
\begin{center}
\caption{The rms radii $R$ and monopole transition matrix elements to the ground state $M({\rm E}0)$ in units of fm and fm$^2$, respectively. $R_{\rm exp.}$ and $M({\rm E}0)_{\rm exp.}$ are the corresponding experimental data.}\label{tab:1}
\begin{tabular}{cccccccccc}
\hline\hline
 & & \multicolumn{2}{c}{$R$} & & \multicolumn{2}{c}{$M({\rm E}0)$} & & $R_{\rm exp.}$ & $M({\rm E}0)_{\rm exp.}$  \\
\hline
 & & SW & MHN & & SW & MHN & & & \\
\hline
$0_1^+$ & & $2.7$ & $2.7$ & &        &       & & $2.71\pm0.02$ &          \\
$0_2^+$ & & $3.0$ & $3.0$ & &  $4.1$ & $3.9$ & &        &  $3.55\pm 0.21$  \\
$0_3^+$ & & $2.9$ & $3.1$ & &  $2.6$ & $2.4$ & &        &  $4.03\pm 0.09$  \\
$0_4^+$ & & $4.0$ & $4.0$ & &  $3.0$ & $2.4$ & &        &  no data \\
$0_5^+$ & & $3.1$ & $3.1$ & &  $3.0$ & $2.6$ & &        &  $3.3\pm0.7$   \\
$0_6^+$ & & $5.0$ & $5.6$ & &  $0.5$ & $1.0$ & &        &  no data \\
\hline\hline
\end{tabular}
\end{center}
\end{table}
We show in TABLE~\ref{tab:1} the calculated rms radii and monopole matrix 
elements to the ground state, together with the corresponding experimental 
values.
The $M({\rm E}0)$ values for the $0_2^+, 0_3^+$, and $0_5^+$ states are 
consistent with the corresponding experimental values. As mentioned above, the structures of the $0_2^+$ and $0_3^+$ states are well established as having the $\alpha + ^{12}$C$(0_1^+)$ and $\alpha + ^{12}$C$(2_1^+)$ 
cluster structures, respectively. These structures of the $0_2^+$ and $0_3^+$ states are confirmed in the present calculation. We also mention that the 
ground state is described as having a shell-model configuration within the 
present framework, the calculated rms value agreeing with the observed 
one ($2.71$ fm). 

On the contrary, the structures of the observed $0_4^+$, $0_5^+$ and $0_6^+$ states in Fig.~\ref{fig:1} have, in the past, not clearly been understood, since they have never been discussed with the previous cluster model calculations~\cite{Suz76,baye2,Kat92}. Although Ref.~\cite{thsr} predicts the $4\alpha$ condensate state around the $4\alpha$ threshold, it is not clear which of those states corresponds to the condensate state. For example, the $0_4^+$ state has been considered as one of the candidates~\cite{wakasa}, and also the $0_5^+$ state with large $M(E0)$ value~\cite{14.01,nupecc}, since the strong monopole transition implies a developed cluster structure ~\cite{monopole}. This ambiguous situation stems from the fact that the $\alpha$-condensate-type wave function given in Ref.~\cite{thsr} can only reproduce the ground state and the $4\alpha$ condensate state, but not the $\alpha+^{12}$C configurations which make up a large part of the $^{16}$O-spectrum up to the $0_6^+$ state. 

 As shown in Fig.~\ref{fig:1}, the present calculation succeeded, for the first time, to reproduce the $0_4^+$, $0_5^+$ and $0_6^+$ states, together with the $0_1^+$, $0_2^+$ and $0_3^+$ states. This puts us in a favorable position to discuss the $4\alpha$ condensate state, expected to exist around the $4\alpha$ threshold.

In Table~\ref{tab:1}, the largest rms 
value of about 5 fm is found for the $0_6^+$ state. Compared with the relatively smaller rms 
radii of the $0_4^+$ and $0_5^+$ states, this large size suggests that 
the $0_6^+$ state may be composed of a weakly interacting gas of $\alpha$ 
particles~\cite{foot} of the condensate type.
\begin{figure}[htbp]
\begin{center}
\includegraphics[scale=0.7]{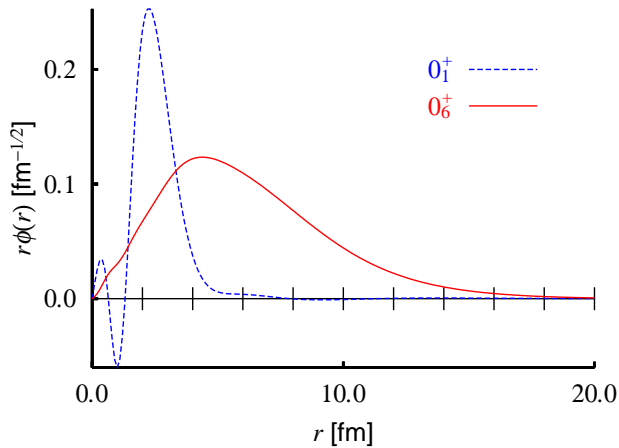}
\end{center}
\caption{(Colors online) The radial parts of single-$\alpha$ orbits with $L=0$ belonging to the largest occupation number, for the ground and $0_6^+$ states with MHN force.}\label{fig:2}
\end{figure}

While a large size is generally necessary for forming an $\alpha$ 
condensate, the best way for its identification is to investigate the single-$\alpha$ orbit and its occupation 
probability, which can be obtained by diagonalizing the one-body ($\alpha$) 
density matrix~\cite{takahashi,matsumura,yamada_ocm} in the following way: 
\begin{equation}
\int d\vc{r}^\prime \rho(\vc{r},\vc{r}^\prime) \phi^L(\vc{r}^\prime) =
\mu^L \phi^L(\vc{r}) ,
\end{equation}
where $\phi^L(\vc{r})$ is the single-$\alpha$ natural orbit with orbital 
angular momentum $L$, and $\mu^L$ is its occupation probability. The one-body density matrix elements 
$\rho(\vc{r},\vc{r}^\prime)$ have been calculated as in ~\cite{matsumura,yamada_ocm}. As a result of the calculation of the $L=0$ case, a large occupation 
probability of $61 \%$ of the lowest $0S$-orbit is found for the $0_6^+$ state, whereas 
the other five $0^+$ states all have appreciably smaller values, at 
most $25 \%$ ($0^+_2$). The corresponding single-$\alpha$ $S$ orbit is 
shown in Fig. \ref{fig:2}. It has a strong spatially extended behaviour 
without any node $(0S)$. This indicates that $\alpha$ particles are 
condensed into the very dilute $0S$ single-$\alpha$ orbit, see also Ref.~\cite{ropke2}. Thus, the $0^+_6$ state clearly has $4\alpha$ condensate character. We should 
note that the orbit is very similar to the single-$\alpha$ orbit of the Hoyle state~\cite{matsumura,yamada_ocm}. We also show in Fig.~\ref{fig:2} the 
single-$\alpha$ orbit for the ground state. It has maximum amplitude at 
around $3$ fm and  oscillations in the interior with two nodal $(2S)$ 
behaviour, due to the Pauli principle and reflecting the shell-model configuration. 
\begin{figure}[htbp]
\begin{center}
\includegraphics[scale=0.7]{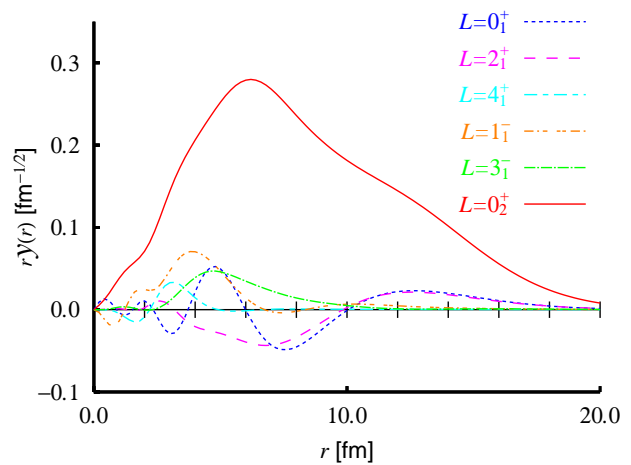}
\end{center}
\caption{(Colors online) $r{\cal Y}(r)$ defined by Eq.~(\ref{eq:rwa}) for the $0_6^+$ state with the MHN force.}\label{fig:3}
\end{figure}

 In order to further analyze the obtained wave functions, we calculate an 
overlap amplitude, which is defined as follows:
\begin{equation}
{\cal Y}(r)= \Big\langle \Big[ \frac{\delta(r^\prime-r)}{r^{\prime 2}}
Y_{L}(\vc{\hat r}^\prime)\Phi_{L}(^{12}{\rm C}) \Big]_{0} \Big| \Psi(0_6^+) 
\Big\rangle. \label{eq:rwa}
\end{equation}
Here, $\Phi_{L}(^{12}{\rm C})$ is the wave function of $^{12}$C, given 
by the $3\alpha$ OCM calculation~\cite{yamada_ocm}, and $r$ is the relative 
distance between the center-of-mass of $^{12}$C and the $\alpha$ particle. 
From this quantity we can see how large is the component in a certain 
$\alpha + ^{12}$C channel which is contained in our wave function (2) for $0_6^+$. The 
amplitudes for the $0_6^+$ state are shown in Fig.~\ref{fig:3}. It only has 
a large amplitude in the $\alpha + ^{12}$C$(0_2^+)$ channel, whereas 
the amplitudes in other channels are much suppressed. The amplitude in the 
Hoyle-state channel has no oscillations and a long tail stretches out to $\sim 20$ fm. This behaviour is very similar to that of the single-$\alpha$ 
orbit of the $0_6^+$ state discussed above. 

 The $\alpha$ decay width constitutes a very important information to identify the $0_6^+$ state from the experimental point of view. It can be estimated, based on the $R$-matrix theory, with the overlap amplitude Eq.~(\ref{eq:rwa})~\cite{r-matrix}. We find that the total $\alpha$ decay width of the $0_6^+$ state is as small as 50 keV (experimental value: 166 keV). This means that the state can be observed as a quasi-stable state. Thus, the width, as well as the excitation energy, are consistent with the observed data. All the characteristics found from our OCM calculation, therefore, indicate that the 6th $0^+$ state shall be identified with the experimental $0_6^+$ state at $15.1$ MeV and that it is the $\alpha$-condensate state of $^{16}$O. 

 Finally we discuss the structures of the $0_4^+$ and $0_5^+$ states. Our present calculations show that the $0_4^+$ and $0_5^+$ states mainly have $\alpha + ^{12}$C$(0_1^+)$ structure with higher nodal behaviour and 
$\alpha + ^{12}$C$(1^-)$ structure, respectively. Further details will be given in a forthcoming extended paper. The calculated width of the $0_4^+$ is $\sim 150$ keV, which is much larger than that found for the $0_5^+$ state $\sim 50$ keV. Both are qualitatively consistent with the corresponding experimental data, $600$ keV and $185$ keV, respectively. The reason why the width of the $0_4^+$ state is larger than that of the $0_5^+$ state, though the $0_4^+$ state has lower excitation energy, is due to the fact that the former has a much larger component of the $\alpha+ ^{12}$C$(0_1^+)$ decay channel, reflecting the characteristic structure of the $0_4^+$ state. The $4\alpha$ condensate state, thus, should not be assigned to the $0_4^+$ or $0_5^+$ state~\cite{4athsr} but very likely to the $0_6^+$ state. 

 In conclusion, the present $4\alpha$ OCM calculation, for the first time, succeeded in describing the structure of the full observed $0^+$ spectrum up to the $0^+_6$ state in $^{16}$O. The $0^+$ spectrum of $^{16}$O up to about 15 MeV is now essentially understood, including the $4\alpha$ condensate state. This is remarkable improvement concerning our knowledge of the structure of $^{16}$O. We found that the $0_6^+$ state above the $4\alpha$ threshold has a very large rms radius of about $5$ fm and a rather strong concentration of $\alpha$ particles $61 \%$ on a spatially extended single-$\alpha$ $0S$ orbit. The wave function was shown to have a large $\alpha + ^{12}$C amplitude only for $^{12}$C (Hoyle state). These results are strong evidence of the newly found $0_6^+$ state as being the $4\alpha$ condensate state, i.e. the analog to the Hoyle state in $^{12}$C. Further experimental information is very much 
requested to confirm the existence of this novel state. Also independent theoretical calculations are strongly needed for confirmation of our results.
 
%\section*{Acknowledgements}

 One of the authors (Y. F.) acknowledges financial assistance from the Special 
Postdoctoral Researchers Program of RIKEN. Numerical calculations were 
performed on the computer facility at RCNP. Correspondence and discussions 
concerning the $^{16}$O spectrum with A. Richter are very much appreciated.


\begin{thebibliography}{99}
%\vspace{-2mm}
\bibitem{tang}
K. Wildermuth and Y. C. Tang, {\it A Unified Theory of the Nucleus} (Vieweg, Braunschweig, 1977).
\bibitem{carbon}
K. Ikeda {\it et al}., Prog. Theor. Phys. Suppl. No. 68, 1 (1980).
\bibitem{roepke}
G. R\"opke {\it et al}., Phys. Rev. Lett. {\bf 80}, 3177 (1998).
%\bibitem{beyer}
%M. Beyer {\it et al}., Phys. Lett. B {\bf 488}, 247 (2000). 
\bibitem{stringari}
F. Dalfovo {\it et al}., Rev. Mod. Phys. {\bf 71}, 463 (1999).
\bibitem{lecheminant}
A. S. Stepanenko {\it et al}., arXiv: cond-mat/9901317; B. Doucot {\it et al}., Phys. Rev. Lett. {\bf 88}, 227005 (2002); H. Kamei {\it et al}., J. Phys. Soc. Jpn. {\bf 74}, 1911 (2005); S. Capponi {\it et al}., Phys. Rev. A {\bf 77}, 013624 (2008).
\bibitem{ring}
P. Ring, and P. Schuck, {\it The Nuclear Many-Body Problem} (Springer-Verlag, Berlin, 1980).
\bibitem{horiuchi_ocm}
H. Horiuchi, Prog. Theor. Phys. {\bf 51}, 1266 (1974);  {\bf 53}, 447 (1975).
\bibitem{kamimura}
Y. Fukushima {\it et al}., Suppl. of J. Phys. Soc. Japan, {\bf 44}, 225 (1978); M. Kamimura, Nucl. Phys. A {\bf 351}, 456 (1981). 
\bibitem{uegaki}
E. Uegaki {\it et al}., Prog. Theor. Phys. {\bf 57}, 1262 (1977); E. Uegaki {\it et al}., Prog. Theor. Phys. {\bf 59}, 1031 (1978); {\bf 62}, 1621 (1979).
\bibitem{baye}
P. Descouvemont {\it et al}., Phys. Rev. C {\bf 36}, 54 (1987).
\bibitem{thsr}
A.~Tohsaki {\it et al}., Phys. Rev. Let. {\bf 87}, 192501 (2001).
\bibitem{funaki1}
Y.~Funaki {\it et al.}, Phys. Rev. C {\bf 67}, 051306(R) (2003).
\bibitem{neff}
M. Chernykh {\it et al}., Phys. Rev. Lett. {\bf 98}, 032501(2007).
\bibitem{matsumura}
H. Matsumura {\it et al}., Nucl. Phys. A {\bf 739}, 238 (2004).
\bibitem{yamada_ocm}
T.~Yamada {\it et al}., Euro. Phys. J. A {\bf 26}, 185 (2005).
%\bibitem{yamada_multi}
%T. Yamada {\it et al}., Phys. Rev. C {\bf 69}, 024309 (2004).
%\bibitem{funaki_res}
%Y. Funaki {\it et al}., Eur. Phys. J. A {\bf 24}, 321 (2005); {\bf 28}, 259 (2006).
%\bibitem{nocore}
%B. R. Barrett {\it et al}., Nucl. Phys. News, {\bf 13}, 17 (2003).
%\bibitem{koka}
%Tz. Kokalova {\it et al}., Eur. Phys. J A {\bf 23}, 19 (2005); Tz. Kokalova {\i%t et al}., Phys. Rev. Lett. {\bf 96}, 192502 (2006).
%\bibitem{kawabata}
%T. Kawabata {\it et al}., Phys. Lett. B {\bf 646}, 6 (2007).
%\bibitem{freer}
%M. Freer {\it et al}., Phys. Rev. C {\bf 71}, 047305 (2005); {\bf 76}, 034320 (2007).%
%\bibitem{ohkubo}
%S. Ohkubo {\it et al}., Phys. Rev. C {\bf 70}, 041602(R) (2004).
%\bibitem{takashina}
%M. Takashina {\it et al}., Phys. Rev. C {\bf 74}, 054606 (2006).
%\bibitem{enyo}
%Y. Kanada-En'yo, Phys. Rev. C {\bf 75}, 024302 (2007).
\bibitem{Suz76}
Y. Suzuki, Prog. Theor. Phys. {\bf 55}, 1751 (1976); {\bf 56}, 111 (1976).
\bibitem{baye2}
M. Libert-Heinemann {\it et al}., Nucl. Phys. A {\bf 339}, 429 (1980).
\bibitem{Hor68}
H.~Horiuchi {\it et al}., Prog. Theor. Phys. {\bf 40}, 277 (1968).
\bibitem{Kat92}
K.~Fukatsu {\it et al}., Prog. Theor. Phys. {\bf 87}, 151 (1992).
\bibitem{saitoh}
S. Saito, Prog. Theor. Phys. {\bf 40}, 893 (1968); {\bf 41}, 705 (1969); Prog. Theor. Phys. Suppl. No. 62, 11 (1977).
\bibitem{kukulin}
V. I. Kukulin {\it et al}., Nucl. Phys. A {\bf 417}, 128 (1984).
\bibitem{mhn}
A. Hasegawa {\it et al}., Prog. Theor. Phys. {\bf 45}, 1786 (1971); F. Tanabe {\it et al}., ibid. {\bf 53}, 677 (1975).
\bibitem{sw}
E. W. Schmid {\it et al}., Nucl. Phys. {\bf 26}, 463 (1961).
\bibitem{kurokawa}
C. Kurokawa {\it et al}., Phys. Rev. C {\bf 71}, 021301 (2005); Nucl. Phys. A {\bf 792}, 87 (2007).
\bibitem{GEM}
M. Kamimura, Phys. Rev. A {\bf 38}, 621 (1988); E. Hiyama {\it et al}., Prog. Part. Nucl. Phys. {\bf 51}, 223 (2003).
%\bibitem{ikeda}
%K. Ikeda, N. Takigawa, and H. Horiuchi, Prog. Theor. Phys. Suppl. Extra Number, 464 (1968).
\bibitem{ajze}
F. Ajzenberg-Selove, Nucl. Phys. A {\bf 46}, 1 (1986).
\bibitem{wakasa}
T.~Wakasa {\it et al.}, Phys. Lett. B {\bf 653}, 173 (2007).
\bibitem{14.01}
M. Stroetzel {\it et al}., Phys. Lett. {\bf 29B}, 306 (1969).
\bibitem{nupecc}
Y.~Funaki {\it et al}., Nucl. Phys. News, {\bf 17}(04), 11 (2007).
\bibitem{monopole}
T. Yamada {\it et al}., arXiv:nucl/th0703045.
\bibitem{foot}
 The value of the density of this state may be too low, 
since the extension of states around and above threshold depend very 
sensitively on their precise position.
\bibitem{takahashi}
Y. Suzuki {\it et al}., Phys. Rev. C {\bf 65}, 064318, (2002).
\bibitem{ropke2}
Y. Funaki {\it et al}., arXive:nucl/th-0801.3131.
\bibitem{r-matrix}
A. M. Lane {\it et al}., Rev. Mod. Phys. {\bf 30}, 257 (1958).
\bibitem{4athsr}
The reason why we previously~\cite{nupecc} assigned the $0_4^+$ and $0_5^+$ states as being candidates for $\alpha$-condensation will be explained elsewhere. 
\end{thebibliography}
\end{document}